\documentclass[prl,amsfonts,twocolumn,showpacs,floatfix]{revtex4}
\usepackage{amsmath}
\usepackage{amsfonts}
\usepackage{amssymb}
\usepackage[dvips]{graphicx}%
  
\begin{document}
\title{Mechanism of hopping transport in disordered Mott insulators}
\author{S. Nakatsuji,$^{1}$ V. Dobrosavljevi\'c,$^{2}$ D. Tanaskovi\'c,$^{2}$ M.
Minakata,$^1$ H. Fukazawa,$^1$ and Y. Maeno $^{1,3}$}
\affiliation{$^1$ Department of Physics, Kyoto University, Kyoto 606-8502, Japan\\
$^2$ National High Magnetic Field Laboratory (NHMFL),
Tallahassee, Florida 32310\\
$^3$ International Innovation Center, Kyoto University, Kyoto
606-8501, Japan}
\date{\today}
\begin{abstract}
By using a combination of detailed experimental studies and simple theoretical
arguments, we identify a novel mechanism characterizing the hopping transport
in the Mott insulating phase of Ca$_{2-x}$Sr$_x$RuO$_4$ near the metal-insulator
transition. The hopping exponent $\alpha$ shows a systematic evolution
from a value of $\alpha=1/2$ deeper in the insulator to the conventional Mott value $\alpha=1/3$ closer to the transition. This behavior, which we
argue to be a universal feature of disordered Mott systems close to the
metal-insulator transition, is shown to reflect the gradual emergence of
disorder-induced localized electronic states populating the Mott-Hubbard gap.
\end{abstract}
\maketitle

The interplay of electron correlation and disorder near metal-insulator
transitions represents one of the most fundamental, yet least
understood problems in contemporary condensed matter physics.
These issues are of particular importance in strongly correlated
materials.
For example, recent studies
on manganites have revealed the emergence of coexisting clusters
near the M-I transition, which plays a crucial role in the
mechanism for colossal magnetoresistance \cite{TokuraCMR}. Similar
inhomogeneities have been observed in underdoped high-$T_{\rm c}$ cuprates, where spin glass and stripe formation feature disorder effects near the M-I transition \cite{cuprates}. Despite extensive experimental efforts \cite{TokuraCMR,cuprates,SarmaPRL},
however, the effects of the intrinsic randomness on the
doping/substitution in Mott transition systems remain largely an
open problem, especially regarding the process of closing the
Mott-Hubbard gap. 

Among the Mott transition systems, the single layered ruthenate Ca$_{2-x}$Sr$_x$RuO$_4$ provides a rare
case of the M-I transition that is highly susceptible to 
chemical pressure because of strong lattice-orbital coupling
\cite{NakatsujiPRL, FriedtCSRO,NakatsujiPRL2}. While
one end member, Sr$_2$RuO$_4$ is a spin-triplet superconductor
\cite{MaenoNature}, complete substitution of isovalent Ca changes
the superconductor into the Mott insulator Ca$_2$RuO$_4$ by
introducing distortions to RuO$_6$ octahedra
\cite{NakatsujiCRO,BradenCRO,AlexanderCRO,FukazawaCRO}. The
lattice deformation in both pure and Sr substituted Ca$_2$RuO$_4$
stabilizes the insulating state by inducing orbital polarization
(orbital order) that creates a half-filled configuration near
$\varepsilon_{\rm F}$
\cite{MizokawaCRO,LeePRL,JungPRL,Anisimov,FangPRB}. In addition,
the results by both Raman and optical conductivity measurements indicate that the superexchange coupling $J$, and the effective
bandwidth ($\propto \sqrt J$) in the insulating region are relatively insensitive to Sr content \cite{LeePRL,RhoPRB}.
Instead, the Sr substitution mainly controls the orbital
polarization; the strong orbital polarization becomes gradually relaxed with
substitution, but suddenly vanishes at the M-I transition to stabilize the metallic state \cite{MizokawaCRO,LeePRL,JungPRL,Anisimov,FangPRB}.

This system is particularly suitable for the study of disorder
effects in closing of the Mott-Hubbard gap because the isovalent
Sr substitution provides statistical randomness 
without altering the key parameters: the net carrier density and
the effective bandwidth. Furthermore, the variation in other
parameters that usually complicates the analysis (crystal
structure, phonon dispersions, etc.) should be small since a
moderate substitution of Sr ($< 10$ \%) is sufficient to suppress
the Mott insulating state.

This paper presents the study of the effects of intrinsic
randomness due to Sr substitution on the M-I transition in
Ca$_{2-x}$Sr$_x$RuO$_4$. In this system, a well-defined Mott
insulator appears at low $T$s by the first order M-I
transition in $0 \leq x < 0.2$. In the insulating
phase, a systematic change in the transport and thermodynamic
properties is observed, which elucidates the evolution of strongly
localized states in closing the Mott-Hubbard gap. Furthermore, 
these localized states most likely reflect the
tendency to spontaneously form inhomogeneities due to nano-scale
phase separation in the vicinity of the Mott transition, as reported in 
other oxides \cite{TokuraCMR,cuprates}. 
Our analysis reveals that this situation gives rise to a novel
mechanism for hopping transport, unique to insulating
phases of strongly correlated systems with moderate disorder.

\paragraph*{Experiment.}
Single crystals of Ca$_{2-x}$Sr$_x$RuO$_4$ were
prepared using a floating zone method \cite{NakatsujiJSSC}. The molten zone was well stirred during the growth to mix Ca/Sr homogeneously. 
X-ray powder diffraction show single phase samples without chemical segregation.
Thus, Ca/Sr mixture provides a statistical disorder that prohibits classical percolation effects.
Resistivity was measured by a standard four-probe dc technique. 
Magnetization was measured with a com1ercial SQUID magnetometer. 
Specific heat was measured by a thermal relaxation method.

\begin{figure}[h]
\centering
\includegraphics[width=\linewidth,clip,trim=0cm 0cm 3cm 0cm]{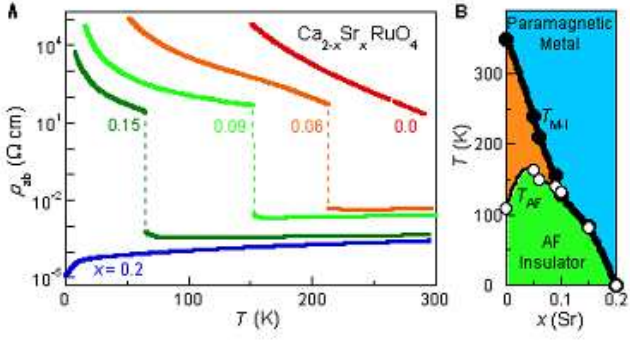}
%\vspace{-1.8in}
\caption{(A) $T$ dependence of the in-plane
resistivity ${\rho_{\rm ab} (T)}$ measured on cooling for Ca$_{2-x}$Sr$_x$RuO$_4$ . Vertical broken lines are guides to the
eye. (B) Phase diagram for the region $0 \leq x \leq 0.2$ with three different phases: paramagnetic metal (blue), paramagnetic insulator (red), and antiferromagnetic insulator (green). $T_{\rm M-I}$ and $T_{\rm AF}$ are the M-I and AF transition temperatures determined on cooling. Thick and thin lines indicate
the first and second order transitions.}
\label{TdepRho}
\end{figure}

Figure 1A presents the temperature dependence of the in-plane
resistivity ${\rho_{\rm ab} (T)}$ measured on cooling . While
Ca$_2$RuO$_4$ ($x$ = 0) is a Mott insulator at low $T$ and
exhibits a M-I transition above 300 K
\cite{NakatsujiCRO,BradenCRO,AlexanderCRO,FukazawaCRO},
slight Sr substitution dramatically decreases the M-I transition temperature
$T_{\rm M-I}$ to below 300 K. The transition is clearly seen by an
abrupt increase of ${\rho_{\rm ab}}$ by a factor of more than
$10^4$. This change at $T_{\rm M-I}$ always involves large thermal
hysteresis, indicating that the M-I transition is first order.
This is due to the first-order structural transition that occurs
simultaneously with the M-I transition \cite{FriedtCSRO}.
Reflecting the quasi-two-dimensional electronic structure, the
anisotropy ${\rho_{\rm c} (T)}/{\rho_{\rm ab} (T)}$ in the insulating
phase reaches the value of the order of $10^3- 10^4$ (not shown).
As illustrated in Fig. 1A and 1B, ${\rho_{\rm ab}}$
in the metallic phase and $T_{\rm M-I}$, both rapidly decrease with $x$, and the system finally becomes fully metallic at $x = 0.2$.

In the insulating phase, an AF ordering appears as in Ca$_2$RuO$_4$.
Figure 2 shows the $T$ dependence of the in-plane component of the
susceptibility $\chi(T)$ measured in a field-cooled sequence.
The distinct increase in $\chi(T)$ at low $T$ arises from canted AF,
as clarified by neutron diffraction measurements \cite{FriedtCSRO}.
The AF transitions are different in nature in the following two regions; in $0 \leq x < 0.1$, the AF ordering occurs at $T < T_{\rm M-I}$ without any hystersis,
while in $0.1 \leq x \leq 0.15$, it coincides with the M-I transition.
In fact, as for $x$ = 0.06 and 0.09, the M-I transition is reflected
in the susceptibility hysteresis in the paramagnetic state
(see the inset I of Fig. 2). 
The decrease of the paramagnetic $\chi$ at $T_{\rm M-I}$
is attributable to the disappearance of Pauli component in the insulating phase.
However, at $x$ = 0.15, a large hysteresis is observed at $T_{\rm
AF}$, indicating that both M-I and AF transitions at $x$ = 0.15
occurs concomitantly with the first order structural transition.

\begin{figure}[h]
\centering
\includegraphics[width=\linewidth,clip,trim=0cm 0cm 9cm 0cm]{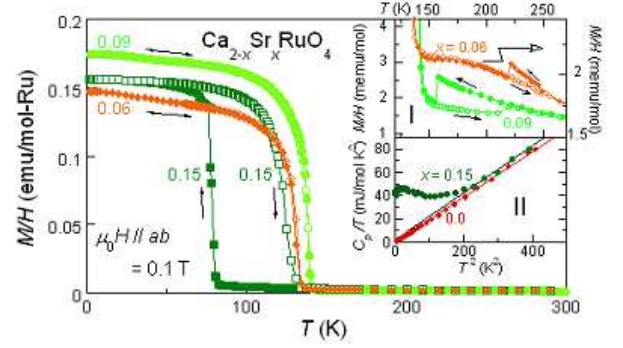}
\vspace{-1in}
\caption{$T$ dependence of the in-plane susceptibility for
Ca$_{2-x}$Sr$_x$RuO$_4$ with $x$ = 0.06, 0.09, and 0.15. All
curves were measured in a field-cooled sequence. The solid and
open symbols indicate the results measured on cooling and heating,
respectively. Inset: (I) thermal hystereses observed in the
paramagnetic phase for $x = 0.06$ and 0.09. (II) $C_P/T$ vs. $T^2$ for $x = 0$ and 0.15. The solid lines
represent linear fit.} \label{inplanesus}
\end{figure}

Here, we note two important facts that confirm the Mott
insulating ground state in $0 \leq x < 0.2$. First,
the AF ordering occurs only in the insulating phase. Second,
$\rho(T)$ does not show any anomaly at $T_{\rm AF}$ for $x =$
0.06 and 0.09.
These facts agree well with the significant feature
of Mott insulators: the separation between charge and spin degrees
of freedom due to a large charge (Mott-Hubbard) gap compared to
low energy spin excitations.

\paragraph*{Observation of hopping transport.} In order to elucidate the
electronic state in the insulating phase, we analyze ${\rho_{\rm
ab}(T)}$ shown in Fig. 1A. The result of Ca$_2$RuO$_4$ ($x$ = 0)
fits well to activation type insulating behavior
\begin{equation}
{\rho_{\rm ab} (T)} = A {\rm exp}(E_{\rm G}/2k_{\rm B} T)
\label{eq:activation}
\end{equation}
below 250 K to the lowest $T$ measured, giving $E_{\rm G} \simeq
4500$ K, consistent with the gap observed by the optical
conductivity measurements \cite{PuchkovPRL,JungPRL}.
Since Ca$_2$RuO$_4$ is a
Mott insulator, $E_{\rm G}$ should give the gap size
between the upper and lower Hubbard bands (UHB and LHB).

Variable-range hopping (VRH) conduction
\begin{equation}
{\rho_{\rm ab} (T)} = A {\rm exp}(T_{\rm 0}/T)^{\alpha}
\label{eq:vrh}
\end{equation}
with $\alpha$ = 1/2 also describes the $T$ dependence over almost the
same $T$ region in agreement with Ref. \cite{AlexanderCRO}.
VRH is usually observed in systems with
strongly localized states near $\varepsilon_{\rm F}$.
However, fitting of the resistivity in a limited region alone does
not give conclusive evidence for the presence of localized
electronic states. In the case of Ca$_2$RuO$_4$, we measured
the specific heat $C_P(T)$ and found the electronic specific heat coefficient
$\gamma$ to be $0 \pm 0.5$ mJ/molK$^2$, as shown in the inset II of Fig. 2.
Therefore, it is most likely that
there are no localized states near $\varepsilon_{\rm F}$ and that
activation-type behavior is the consistent interpretation of the
low $T$ transport in Ca$_2$RuO$_4$.

\begin{figure}[h]
\centering
\includegraphics[width=\linewidth,clip,trim=0cm 0cm 2.5cm 0cm]{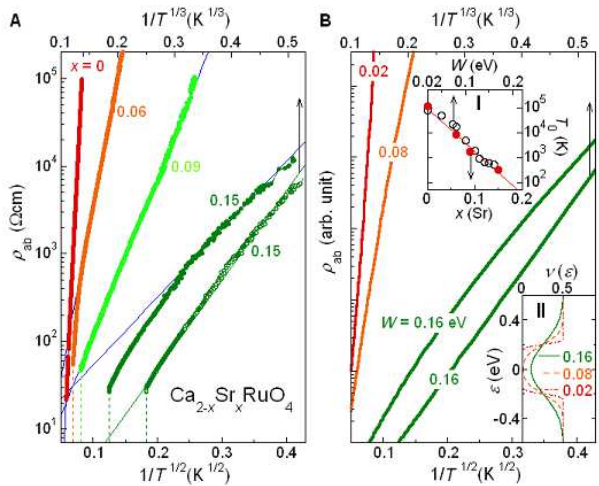}
\vspace{-0.3in}
\caption{ (A) $\ln \rho$ vs. $1/T^{1/2}$ and $1/T^{1/3}$
for the insulating phase of Ca$_{2-x}$Sr$_x$RuO$_4$.
The results for $x \leq 0.09$
are plotted against the lower horizontal axis $1/T^{1/2}$, while
these for $x = 0.15$ are plotted against both $1/T^{1/2}$ and the
upper axis $1/T^{1/3}$. Straight lines indicate a
fitting result to Eq.\ (\ref{eq:vrh}).
(B) Corresponding plot for theoretical predictions.
The results for $W$ = 0.02 and 0.08 eV are plotted against the lower axis $1/T^{1/2}$, while these for $W$ = 0.16 eV are plotted against both $1/T^{1/2}$ and the upper axis $1/T^{1/3}$. 
Inset: (I) $T_{\rm0}$ obtained by fitting of $\rho(T)$ to Eq. (2) with $\alpha = 1/2$ for Ca$_{2-x}$Sr$_x$RuO$_4$ (solid circle) as a function of $x$ (lower axis), and for theory (open circle) as a function of $W$ (upper axis). 
(II) Model band structure used in our calculation. As the disorder increases ($W$ : 0.02 eV $\rightarrow$ 0.16 eV), the band-tails gradually fill-in the Mott gap, producing a soft gap similar to that predicted by Efros and Shklovskii, but on a much larger energy scale.}
\label{lnrho}
\end{figure}

To obtain the systematic estimate of $E_{\rm G}$, we also try to
fit the $\rho(T)$ curve for each $x$ to Eq.\ (\ref{eq:activation}).
However, the well fitted region
becomes rapidly narrower with $x$, indicating that the Sr
substituted region does not obey activated behavior.
In contrast, as in Fig. 3A, the VRH widely describes the insulating
behavior. Equation\
(\ref{eq:vrh}) with $\alpha = 1/2$ describes $\rho_{\rm ab}(T)$
for $x$ = 0.06 and 0.09 over a respectable temperature range: 236
K-150 K ($x$ = 0), 155 K-46 K (0.06), 152 K-14.0 K (0.09), and 21
K-7.3 K (0.15). It should be noted that $\rho_{\rm ab}(T)$ obeys
VRH with $\alpha = 1/2$ over a decade of $T$ for $x$ = 0.09 and
over a comparable range for $x$ = 0.06. The $x$ dependence of the
fitting parameter $T_0$ is given in the inset I of Fig. 3B, showing
exponential decrease with $x$.

For the highest Sr concentration  $x$ = 0.15, $\rho_{\rm ab}(T)$
does not follow Eq.\ (\ref{eq:vrh}) well with $\alpha = 1/2$, but
better fits are obtained with a smaller value of the exponent
$\alpha$ = 1/3 over a decade of $T$ (5 K - 65 K), as shown in
Fig. 3A. The fit gives $T_{\rm 0}$ = $8.8 \times 10^3$ K. This
value of the hopping exponent $\alpha$ is consistent with
two-dimensional (2D) transport, within the Mott VRH picture
\cite{MottNonCryst}. Because our system has essentially a quasi 2D
electronic structure with large anisotropy $\rho_{\rm c}/\rho_{\rm
ab} \approx 10^3$ in the insulating phase, 2D hopping seems a
natural choice.

\paragraph*{Origin of hopping transport.}
Generally, in the process of carrier doping in
conventional semiconductors such as Si and Ge,
the evolution of the transport behavior from simple activation to
Efros-Shklovskii (ES) VRH (Eq.\ (\ref{eq:vrh}) with $\alpha =
1/2$) and finally to Mott VRH has been observed, as
the localized states fill in the band gap \cite{Shklovskii-Efros}.
The Mott VRH, observed at $x$ = 0.15 for a decade of $T$, usually appears
 when the system has
a localized band with a nearly constant density of states (DOS)
 $\nu (\varepsilon )$
around $\varepsilon_{\rm F}$. The low temperature $C_P/T$ for $x$
= 0.15 is well described by $ \gamma + \beta T^{\rm 2}$ with the
same Debye temperature ${\mit\Theta}_{\rm D} = 410$ K as
Ca$_2$RuO$_4$ (411 K) and Sr$_2$RuO$_4$ (410 K) (Inset II of Fig. 2).
The electronic contribution $\gamma$ is about $1.5 \pm 0.5$
mJ/molK$^2$. The extra entropy release of 0.36 J/mol K due to the
increase in $C_P/T$ below 14 K is much smaller than $R$ln3 = 9.13
J/mol K expected for $S = 1$ moment at each Ru site, and is
attributable to the freezing of the uncorrelated spins induced by
disorder. Therefore, the $\gamma$ value above should give an
appropriate estimate of $\nu (\varepsilon_{\rm F})$ at $T > 14$ K.
For 2D Mott VRH, the localization length $\xi$ can be estimated
using the formula $k_{\rm B}T_0 \simeq 8.6/\nu (\varepsilon_{\rm F})
\xi^{2}d$, where $d$ is interlayer spacing. Our results yield a
reasonable value of $\xi \approx 25$ \AA, that is several times
the Ru-Ru interatomic spacing ($\approx 3.8$ \AA).

In contrast, Efros-Shklovskii VRH usually appears when the
long-range Coulomb interaction between localized electrons is
important, forming the so-called Coulomb gap (CG) with $\nu
(\varepsilon) \propto |\varepsilon - \varepsilon_{\rm F}|$
in 2D \cite{Shklovskii-Efros}.
This leads to VRH with $\alpha = 1/2$ and $k_{\rm B}T_0 \simeq
e^2/\kappa \xi$, where $\kappa $ is a dielectric constant.
However, $ T_0$ observed in our system (Inset I of Fig. 3B) is too
 large for long range Coulomb interaction.
In fact, assuming $\kappa \approx 100$ as in
 Ca$_2$RuO$_4$ \cite{AlexanderCRO}, we obtain $\xi$ of the order of
 0.03 \AA for $x$ = 0.06, which is not physically meaningful. Correspondingly,
 the energy scale of fitting regions ($T \leq 150$ K) is also much too
 high in comparison with the ordinary size of CG (10-100 K) for
 semiconductors \cite{Shklovskii-Efros}.
It is therefore unlikely that the emergence of the Coulomb gap is the physical origin of the
apparent ES-like behavior.
%{Nevertheless, $\alpha= 1/2$ is clearly observed for almost a decade of
% $T$ at $x$ =0.06 and 0.09, establishing the insufficiency of the Mott
% VRH mechanism for these materials.}

What could then lie at the origin of the observed VRH with
$\alpha\approx 1/2$? To answer this question, we note that the
gap-type structure in the DOS is essential for the exponent $\alpha$ = 1/2;
The standard Mott VRH picture assumes an energy-independent DOS, leading
 to a small hopping exponent $\alpha$ = 1/3 in 2D, while the formation of
the CG gives a strongly energy-dependent DOS, which in turn produces a
 larger hopping exponent $\alpha$ = 1/2. However, we have seen that the
 CG is too small to account for our observations. Instead, given the large energy
 scale of both the fitting region and $T_0$, the energy gap has to be $0.1-1$ eV,
 which actually fits well in the range of the Mott gap.
Since the insulating phase of our system is a Mott insulator, we
propose that the disorder-modified Mott
gap lies at the origin of the VRH with $\alpha\approx 1/2$, as a
new mechanism distinct from the ES scenario.

Materials near a first-order transition separating two
competing ground states are generally fragile against phase
separation; moderate randomness can create coexisting clusters of
competing ordered states \cite{TokuraCMR}. In our case, the
first order M-I transition separates the Mott insulating phase and
the paramagnetic metallic phase. The statistical distribution of
Ca/Sr inevitably creates (although small) the locally
Sr-rich region with a broader $local$ bandwidth.
This statistical distribution of the local bandwidth should create
extended tails of both UHB and LHB at their edges, 
which are easily localized. As the system
approaches the metallic phase, the band tails extend deeper into 
the Fermi level region by enhancing their DOS. Finally, for large enough $x$(Sr), these tails are
expected to overlap and produce strongly localized states near
$\varepsilon_{\rm F}$. As the intrinsic disorder steadily
increases, we thus expect the behavior to gradually cross over
from activated, to ES-like VRH, and finally to Mott-like VRH. This is
exactly what is observed in our experiment.

\paragraph*{Theory.}
To put these physical ideas on a more rigorous
basis, we now present a simple theoretical model for hopping
transport in disordered Mott insulators. First, we generalize the
arguments of Mott \cite{MottNonCryst} and Efros-Shklovskii
\cite{Shklovskii-Efros} to a situation where the DOS has an
arbitrary energy dependence. According to these approaches, the
transition probability between two localized states is
\begin{equation}
\sigma=\sigma_{o}\exp\{-2R/\xi-\varepsilon_{o}(R)/k_{\rm B} T\},\label{sigma}%
\end{equation}
where $\sigma_{o}$ is an ``attempt frequency," $R$ is the distance
between the two sites, $\varepsilon_{o}(R)$ is their energy
difference, and $\xi$ is the localization length. The further the
sites are from each other, the smaller the wave-function overlap between the two
states. However, sites with small energy difference are typically
more distant from each other. The lowest energy $\varepsilon_{o}(R)$ of
the accessible state within a radius $R$ from a given site
is implicitly given (in 2D) by the solution of
\begin{equation} 1= \pi R^2 H(\varepsilon_o (R) ),\end{equation} where
$ H(\varepsilon)=
\int_{o}^{\varepsilon}\nu(\varepsilon^{\prime})d\varepsilon
^{\prime}$. At any given temperature, the most probable hopping
distance $R$ is determined by minimizing the exponent of Eq.
(\ref{sigma}), giving
\begin{equation}
\frac{1}{\xi}
 =\frac{\left(\pi H(\varepsilon_{o})\right)  ^{3/2}%
}{\pi\nu(\varepsilon_{o})k_{\rm B} T}.\label{epsilont}%
\end{equation}

The solution of these equations determines $R(T)$ and thus
$\varepsilon_{o}(T)$, which should be substituted in Eq.
(\ref{sigma}), to obtain the desired $T$-dependent
resistivity $\rho (T)\sim 1/\sigma$. For the special cases of
constant or linear forms for $\nu(\varepsilon)$, these equations
can be solved analytically, reproducing the respective Mott
($\alpha$ = 1/3) or Efros-Shklovskii ($\alpha$ = 1/2) hopping
laws. In our case, $\nu(\varepsilon)$ assumes a more complex
form, and a numerical solution is necessary.
Given our fortuitous situation
where the Sr substitution does not make significant change in
net carrier density nor effective bandwidth,
we chose a simple model for $\nu(\varepsilon)$.
Specifically, we take the UHB and LHB to have constant DOS and bandwidth
 of 1 eV, with a gap of 0.4 eV, and $\xi = 28$ \AA, which 
corresponds to our material. The effect of disorder is
described by introducing gaussian band-tails of width $W$, which
tend to gradually fill-in the gap, as shown in the inset II of Fig.~3B.
We examine the evolution of $\rho (T)$ and the fitting parameter $T_0$ to
VRH with the exponent $\alpha$ = 1/2, as a function of disorder $W$.
The theoretical results are presented in Fig. 3B for $\rho (T)$
 in the same fashion as the experimental ones in Fig. 3A, and in the inset I 
of Fig. 3B for $T_0(W)$.
 The theory is able to reproduce all the qualitative and even some
 quantitative aspect of the experimental data for both $\rho (T)$ and
 $T_0$, giving strong support to the proposed physical picture.

In summary, detailed analysis of the Mott insulating phase in
Ca$_{2-x}$Sr$_x$RuO$_4$ reveals that the Sr substitution to
Ca$_2$RuO$_4$ changes transport from the activation type to the
variable-range-hopping. The moderate randomness intrinsic to the Sr
substitution leads to the formation of strongly localized states
around the closing of the Mott-Hubbard gap. As a result, hopping
transport emerges of the form similar to that predicted by Efros
and Shklovskii, but with a distinctly different physical origin,
and on a distinctly different energy scale.

The authors acknowledge T. Ishiguro for his support, and S. McCall,
Z. Ovadyahy for comments. This work was supported in part by Grants-in-Aid
for Scientific Research from JSPS and for the 21st Century COE ``Center for Diversity and Universality in Physics" from MEXT of Japan, and by the Sumitomo Foundation. Work at FSU was supported by the NSF grant DMR-0234215.

\end{document}